\documentclass[twocolumn,secnumarabic,amssymb, nobibnotes, aps, prd]{revtex4}
\usepackage{amsmath} \usepackage{amssymb}
\usepackage{graphicx} 
\usepackage{epstopdf}
\usepackage{amsmath}
\usepackage{amsmath,amssymb,amsthm,amsfonts,mathrsfs,bm,verbatim}
\usepackage{graphicx,subfigure}

\newcommand{\bea}{\begin{eqnarray}}
\newcommand{\eea}{\end{eqnarray}}
\def\nn{\nonumber}

\begin{document}

\title{Analytic study of superradiant stability of Kerr-Newman black holes under charged massive scalar perturbation}%

\author{Jun-Huai Xu}
\affiliation{School of Information and Optoelectronic Science and Technology,South China Normal University,Guangzhou 510006,China}
\author{Zi-Han Zheng}
\affiliation{Guangdong Provincial Key Laboratory of Quantum Engineering and Quantum Materials,\\ School of Physics and Telecommunication Engineering,\\
South China Normal University,Guangzhou 510006,China}
\author{Ming-Jian Luo}
\affiliation{Guangdong Provincial Key Laboratory of Quantum Engineering and Quantum Materials,\\ School of Physics and Telecommunication Engineering,\\
South China Normal University,Guangzhou 510006,China}
\author{Jia-Hui Huang}
\email{huangjh@m.scnu.edu.cn}
\affiliation{Guangdong Provincial Key Laboratory of Nuclear Science, Institute of quantum matter,South China Normal University, Guangzhou 510006, China
}
\affiliation{Guangdong-Hong Kong Joint Laboratory of Quantum Matter, Southern Nuclear Science Computing Center, South China Normal University, Guangzhou 510006, China}

\begin{abstract}
The superradiant stability of a Kerr-Newman black hole and charged massive scalar perturbation is investigated. We treat the black hole as a background
geometry and study the equation of motion of the scalar perturbation. From the radial equation of motion, we derive the effective potential experienced by
the scalar perturbation. By a careful analysis of this effective potential, it is found that
when the inner and outer horizons of Kerr-Newman black hole satisfy $\frac{r_-}{r_+}\leqslant\frac{1}{3}$  and the charge-to-mass ratios of scalar perturbation and black hole
satisfy $ \frac{q}{\mu }\frac{Q}{ M}>1 $, the Kerr-Newman black hole and scalar perturbation system is superradiantly stable.
\end{abstract}

\maketitle

\section{Introduction}
Black hole physics has been studied extensively for a long time in general relativity. Recently, due to the first image of the black hole at the center
of the galaxy M87 \cite{Akiyama:2019bqs,Akiyama:2019cqa}, the study of the motion of electromagnetic fields or other matter fields in the background of black holes has been attracted much attention\cite{Himwich:2020msm,Johnson:2019ljv,Gralla:2019drh,Gralla:2019xty}.

Superradiance is an interesting phenomenon which can be used to extract electromagnetic or rotating energy from black holes by scattering off the black holes with bosonic matter fields (for a comprehensive review, see \cite{Brito:2015oca}). For a charged rotating black hole,
an impinging charged bosonic wave will be amplified by the black hole when the wave frequency $\omega$ satisfies
\bea\label{src}
\omega< m\varOmega_H+q\varPhi_H,
\eea
where $q$ and $m $ are the charge and azimuthal quantum number of the incoming wave, $\varOmega_H$ is the angular velocity of black hole horizon and $\varPhi_H$ is the electromagnetic potential of the black hole horizon. If there is a mirror-like mechanism between the black hole horizon and space infinity, the amplified superradiant modes may be scattered back and forth and grows exponentially, which is called a black hole bomb mechanism proposed long time ago\cite{PTbomb} and may lead to the superradiant instability of the background black hole geometry\cite{Cardoso:2004nk,Herdeiro:2013pia,Degollado:2013bha}.

Superradiant (in)stability of various kinds of black holes have been studied extensively in the literature. Charged Reissner-Nordstrom (RN) black holes were proved superradiantly stable against charged massive scalar perturbation \cite{Hod:2013eea,Huang:2015jza,Hod:2015hza,DiMenza:2014vpa}. The argument is as follows: when superradiant modes exist for a charged massive scalar perturbation in a RN black hole background, there is no effective trapping potential/mirror outside the black hole horizon, which reflects the superradiant modes back and forth\cite{Li:2014gfg,Sanchis-Gual:2015lje,Fierro:2017fky,Li:2014fna,Li:2015mqa}. Rotating Kerr black holes may also be superradiantly unstable under massive scalar perturbation, where the mass of the scalar provides a natural mirror\cite{Strafuss:2004qc,Konoplya:2006br,Cardoso:2011xi,Dolan:2012yt,Hod:2012zza,Aliev:2014aba,Hod:2016iri,Degollado:2018ypf,Huang:2019xbu,Hod:2014pza}. If  a mirror-like boundary condition is imposed outside the black hole horizon or if the RN/Kerr black hole is asymptotically non-flat, the black hole is superradiantly unstable in certain parameter spaces\cite{Wang:2014eha,Bosch:2016vcp,Huang:2016zoz,Gonzalez:2017shu,Zhu:2014sya}. Besides scalar perturbations, a massive vector perturbation on a rotating Kerr black hole has also been discussed\cite{East:2017ovw,East:2017mrj}.

The superradiant instability properties of charged and rotating Kerr-Newman (KN) black holes under charged massive scalar perturbation were also discussed in literature\cite{Furuhashi:2004jk,Hod:2016bas,Huang:2016qnk,Huang:2018qdl,Chowdhury:2019ptb}. In \cite{Furuhashi:2004jk}, the growth rate of unstable modes of the massive scalar field was obtained for $\mu M\leq 1$, where $\mu$ is the proper mass of the scalar field and $M$ is the mass of the KN black hole. Maximal growth rates were also found for some chosen parameters. In \cite{Hod:2016bas}, complex resonance spectrum of a charged massive scalar field in a near extremal KN black hole spacetime was studied. It was shown that the superradiant instability growth rates of the scalar fields are characterized by the dimensionless charge-to-mass ratio $q/\mu$, where $q$ is the charge of the scalar field. Due to the electromagnetic interaction between the scalar field and the Kerr-Newman black hole, it was shown that the growth rate of the instability of KN black hole can be larger than that of a scalar field in Kerr spacetime of the same rotation parameter\cite{Huang:2018qdl}.

Although there are some results on instability of KN black hole and massive scalar perturbation system, there is a few study of superradiantly stable regions in the parameter space of the system. In this paper we will use analytic method to find the superradiantly stable parameter space regions for a KN black hole against  charged massive scalar perturbation. The radial equation of motion and effective potential for the scalar perturbation are the main objects in our analysis.  We will find a parameter space where the superradiance condition and bound state condition holds but there is no potential well outside the black hole outer horizon in the effective potential\cite{Huang:2019xbu,Hod:2014pza}. It is noteworthy that unexpected simplicity is found in our analysis of some seemingly complicated coefficients in the effective potential.

The paper is organized as follows. In Section 2, we provide a simple introduction of the KN-black-hole-massive-scalar system and the angular part of the equation of motion. In Section 3, the Schrodinger-like radial equation and the effective potential for the scalar perturbation in KN background are reviewed. Several important asymptotic behaviors of the effective potential and derivative of the effective potential are discussed.  In Section 4,  detailed analysis of the effective  potential is given and the superradiantly stable parameter space regions are obtained. Section 5  is devoted to a summary.

\section{Kerr-Newman black hole and a charged massive scalar perturbation}

In this section, a description of our model and the equation of motion for the scalar perturbation will be given. The physical system consists of a rotating charged Kerr-Newman black hole and a minimally coupled charged  massive scalar perturbation. The metric of the 4-dimensional Kerr-Newman black hole in Boyer-Lindquist coordinates $(t,r,\theta,\phi)$ is (we use natural unit, $G=\hbar=c=1$)
\bea\nn
ds^2&=&-\frac{\varDelta}{\rho ^2}\left( dt-\text{a} \sin ^2\theta d\phi \right) ^2+\frac{\rho ^2}{\varDelta}dr^2\\
&+&\rho ^2d\theta ^2+\frac{\sin ^2\theta}{\rho ^2}\left[ \left( r^2+a^2 \right) d\phi -adt \right] ^2,
\eea
where
\begin{equation}
\rho ^2\equiv r^2+a^2\cos ^2\theta\quad,\quad\varDelta\equiv r^2-2Mr+a^2+Q^2,
\end{equation}
$Q,~M$ are the charge and mass of the KN black hole and $a$ is the black hole angular momentum per unit mass.
The inner and outer horizons of the Kerr-Newman black hole are
\begin{equation}
{{r}_{\pm }}=M\pm \sqrt{{{M}^{2}}-{{a}^{2}}-{{Q}^{2}}},
\end{equation}
and they satisfy the following obvious relations
\begin{equation}
{{r}_{+}}+{{r}_{-}}=2M,~~{{r}_{+}}{{r}_{-}}={{a}^{2}}+{{Q}^{2}}.
\end{equation}
The background electromagnetic potential is
\begin{equation}
A_{\nu}=\left( -\frac{Qr}{\rho ^2},0,0,\frac{aQr\sin ^2\theta}{\rho ^2} \right).
\end{equation}

The equation of motion for a minimally coupled charged massive scalar perturbation field $\Phi$ is governed by the following covariant Klein-Gordon equation
\begin{equation}
( \nabla ^{\nu}-iqA^{\nu})( \nabla _{\nu}-iqA_{\nu}) \Phi =\mu ^2\Phi,
\end{equation}
where $\nabla ^{\nu}$ is the covariant derivative in the KN background. The above equation is separable and the solution of the above equation with definite frequency can be decomposed as
\begin{equation}
\Phi ( t,r,\theta ,\phi) =\sum_{lm}R_{lm}( r ) S_{lm}( \theta) e^{im\phi}e^{-i\omega t}.
\end{equation}
The radial function $R_{lm}$ satisfy the radial part of the equation of motion (see Eq.\eqref{REoM} below), which is the main equation studied in the paper. The angular function $S_{lm}$ are scalar spheroidal harmonics satisfying the angular part of the equation of motion (see Eq.\eqref{AEoM} below). $l(=0,1,2,...)$ and $m$ are integers, $-l\leq m\leq l$ and $\omega$ is the angular frequency of the scalar perturbation.

The angular part of the equation of motion is an ordinary differential equation which is written as follows\cite{Bardeen:1972fi,Berti:2005gp,Hod:2015cqa},
\bea\label{AEoM}\nn
&&\frac{1}{\sin \theta}\frac{d}{d\theta}( \sin \theta \frac{dS_{lm}}{d\theta})\\
 &&+[ \lambda_{lm}+( \mu ^2-\omega ^2) a^2\sin ^2\theta -\frac{m^2}{\sin ^2\theta} ] S_{lm}=0,
\eea
where $\lambda_{lm}$ are angular eigenvalues. This equation is the standard spheroidal differential equation which is important in many physical problems and has been studied for a long time. The spheroidal functions $S_{lm}$ are called prolate (oblate) for $( \mu ^2-\omega ^2) a^2>0( <0)$. In this paper, we will consider only the prolate case.
There is no explicitly analytic expression for angular eigenvalues $\lambda_{lm}$. Here we choose the following lower bound for this separation constant\cite{Hod:2016iri},
\begin{equation}\label{lamda}
\lambda _{lm}>m^2 -a^2( \mu ^2-\omega ^2).
\end{equation}

The radial part of the Klein-Gordon equation satisfied by $R_{lm}$ is given by
\begin{equation}\label{REoM}
\varDelta \frac{d}{dr}( \varDelta \frac{dR_{lm}}{dr}) +UR_{lm}=0,
\end{equation}
where
\bea\nn
U&=&[\omega( a^2+r^2)-am-qQr] ^2\\
&&+\varDelta [ 2am\omega-\lambda _{lm}-\mu ^2( r^2+a^2)].
\eea

\subsection{Boundary Conditions}
In order to study the superradiant modes of KN black hole under the charged massive scalar perturbation, appropriate boundary conditions should be considered for asymptotic solutions of the radial equation near the horizon and at spatial infinity. Here we use tortoise coordinate to analyse the boundary conditions for the radial function.
Define the tortoise coordinate $r_*$ by following equation
\begin{equation}
\frac{dr_*}{dr}=\frac{r^2+a^2}{\varDelta}.
\end{equation}
The boundary conditions we are interested are  purely ingoing wave near the outer horizon and exponentially decaying wave at spatial infinity. Therefore, the asymptotic solutions of the radial wave function at these two boundaries are chosen as follows
\bea\nn
R_{lm}( r) \sim \begin{cases}
	e^{-i( \omega -\omega _c) r_*}, &r^*\rightarrow -\infty\text{(}r\rightarrow r_+ \text{)}\\
	\frac{e^{-\sqrt{\mu ^2-\omega ^2}r}}{r}, &r^*\rightarrow +\infty\text{(}r\rightarrow +\infty \text{)}.\\
\end{cases}
\eea
It is easy to see that in order to get decaying modes at spatial infinity we need following bound state condition
\begin{equation}\label{bsc}
\omega ^2<\mu^2.
\end{equation}
Here the critical frequency $\omega_c$ is defined as
\begin{equation}
\omega _c=m\varOmega _H+q\varPhi _H,
\end{equation}
where $\varOmega _H=\frac{a}{r_{+}^{2}+a^2}$ is angular velocity of the outer horizon and $\varPhi _H=\frac{Qr_+}{r_{+}^{2}+a^2}$ is  the electric potential of outer horizon.

\section{The radial equation of motion and effective potential}

By defining a new radial wave function
\begin{equation}
\psi _{lm}\equiv \varDelta^{\frac{1}{2}}R_{lm}.
\end{equation}
We can transform the radial equation of motion \eqref{REoM} into a Schrodinger-like wave equation
\begin{equation}
\frac{d^2\Psi _{lm}}{dr^2}+( \omega ^2-V) \Psi _{lm}=0,
\end{equation}
where
\begin{equation}
\omega ^2-V=\frac{U+M^2-a^2-Q^2}{\varDelta ^2},
\end{equation}
and $V$ is the effective potential.

Given the superradiant condition \eqref{src}, i.e. $\omega<\omega_c$, and bound state condition \eqref{bsc}, if there is no trapping potential well outside the outer horizon of the KN black hole, the KN black hole and charged massive scalar perturbation system will be superradiantly stable. In the following, we will analyze the shape of the effective potential $V$ to discuss the nonexistence of a trapping well.

The asymptotic behaviors of the effective potential $V$ near the inner and outer horizons and at spatial infinity are
\begin{equation}
V( r\rightarrow +\infty )
\rightarrow \mu ^2-\frac{2( 2M\omega ^2-qQ\omega -M\mu ^2)}{r}+{\cal O}( \frac{1}{r^2}) ,
\end{equation}
\bea
V( r\rightarrow r_+ ) \rightarrow -\infty,~~
V( r\rightarrow r_-) \rightarrow -\infty.
\eea
From the above equations, we know the effective potential approaches to a constant at spatial infinity and there is at least one extreme between inner and outer horizons.

\subsection{Derivative of the effective potential near spatial infinity }
The asymptotic behaviour of the derivative of the effective potential $V$ at spatial infinity is
\begin{equation}
V'( r\rightarrow +\infty)
\rightarrow -\frac{-4M\omega ^2+2qQ\omega +2M\mu ^2}{r^2}+{\cal O}( \frac{1}{r^3}).
\end{equation}
The no trapping well condition requires that the derivative of the effective potential is negative. This means the following quadratic function $f$ for $\omega$
\begin{equation}
f(\omega)=-4M\omega^2+2Qq\omega+2M\mu^2,
\end{equation}
should be positive. There are two real roots for equation $f(\omega)=0$. One is positive and the other one is negative. The positive real root is
\begin{equation}
\omega_+=\frac{Qq+\sqrt{Q^2q^2+8M^2\mu^2}}{4M}.
\end{equation}
Given the bound state condition \eqref{bsc}, in order to obtain $f(\omega)>0$,  we just need $\omega_+>\mu$.
Consider the following difference
\bea\nn
\omega_+-\mu&=&\frac{Qq+\sqrt{Q^2q^2+8M^2\mu^2}}{4M}-\mu\\
&=&\frac{Qq-4M\mu +\sqrt{Q^2q^2+8M^2\mu^2}}{4M},
\eea
if
\begin{equation}
[\sqrt{Q^2q^2+8M^2\mu^2}]^2-(Qq-4M\mu)^2 > 0,
\end{equation}
i.e.
\begin{equation}\label{mu1}
 \frac{q}{\mu }\frac{Q}{ M}>1,
\end{equation}
we will get $\omega_+-\mu>0$ and $f(\omega)>0$. This is one simple and key result of this paper.

Given the condition \eqref{mu1}, there is no trapping well near the spatial infinity and the black hole may be superradiantly stable.
In the next section, we will go a step further and find the superradiant stable regions of the parameter space where there is only one extreme outside the outer horizon $r_+$ and there is no trapping well from the horizon to spatial infinity.

\section{The superradiant stability analysis}
In this section, the superradiant stable proper parameter space regions will be determined for the KN black hole and charged massive scalar perturbation system.
We will analyze this by considering the roots of the derivative of the effective potential between inner horizon and spatial infinity.
\subsection{The derivative of the effective potential}
We define a new radial coordinate $z$, $z=r-{r_-}$. Then the explicit expression of the derivative of the effective potential $V$ in radial coordinates $z$ and $r$ can be written as follows,
\bea\nonumber
V'( r)&=&\frac{Ar^4+Br^3+Cr^2+Dr+E}{-\varDelta ^3}\\
=V'( z)&=&\frac{A_1z^4+B_1z^3+C_1z^2+D_1z+E_1}{-\varDelta ^3}.
\eea
The relations between the two sets of coefficients are
\bea
A_1&=& A,\\
{B_1} &=& B+4r_-A_1,\\
{C_1} &=& C+3r_-B_1-6r_{-}^{2}A_1,\\
{D_1} &=& D+4r_{-}^{2}A_1-3r_{-}^{2}B_1+2r_-C_1,\\
{E_1} &=& E-r_{-}^{4}A_1+r_{-}^{3}B_1-r_{-}^{2}C_1+r_-D_1.
\eea

There are three coefficients which are important for our following analysis. They are listed as follows
\bea
A_1&=&2qQ\omega +2M\mu ^2-4M\omega ^2,\\\nonumber
C_1&=&-3(r_+-r_-)\lambda_{lm}+12r_-(Q^2-r_-^2-r_+r_-)\omega^2\\\nonumber
&&+6[am(r_-+r_+)-(Q^2-3r_-^2-r_-r_+)qQ]\omega\\\nonumber
&&-3(Q^2r_--Q^2r_+-r_-^3+r_+^2 r_-)\mu^2\\
&&-6amqQ-6q^2Q^2r_-,
\eea
\bea\nn
E_1&=&-4a^2M+4a^2m^2M+4M^3-4MQ^2\\\nn
&& +2amqQ( a^2+Q^2) +4a^2M( -a^2-Q^2) \mu ^2\\\nn
&& +2a^2M( a^2+Q^2) \mu ^2+4M( -a^2-Q^2) \lambda _{lm}\\\nn
&& +2M( a^2+Q^2) \lambda _{lm}+4a^2r_--4a^2m^2r_-\\\nn
&& +4amMqQr_-+2a^2q^2Q^2r_-+4a^2M^2\mu ^2r_-\\\nn
&& -2a^2Q^2\mu ^2r_-+4M^2( -1+\lambda _{lm}) r_-\\\nn
&& +2a^2\lambda _{lm}r_-+2Q^2[ 2+Q^2( q-\mu)( q+\mu) \\\nn
&& +\lambda _{lm}] r_--6amqQr_{-}^{2}-6M( -Q^2\mu ^2\\\nn
&& +\lambda _{lm}) r_{-}^{2}-4M^2\mu ^2r_{-}^{3}-2Q^2( q^2+\mu ^2) r_{-}^{3}\\\nn
&& +2\lambda _{lm}r_{-}^{3}+2M\mu ^2r_{-}^{4}+\omega ^2( 4a^4M+4a^2Q^2r_-\\\nn
&& +4Q^2r_{-}^{3}-4Mr_{-}^{4}) +\omega[ -8a^3mM\\\nn
&& -8amM( -a^2-Q^2) -4amM( a^2+Q^2) \\\nn
&& -2a^2qQ( a^2+Q^2)-4a^2MqQr_-\\\nn
&& -8am( M^2+Q^2) r_-+12amMr_{-}^{2}-6qQ^3r_{-}^{2}\\
&& +4MqQr_{-}^{3}+2qQr_{-}^{4}].
\eea

We use $f_1 (z)$ to denote the numerator of the derivative of the effective potential $V'(z)$, which is a quartic polynomial of $z$. We can study whether there is an trapping well outside the horizon by analyzing the property of the roots of the equation $f_1(z) = 0$. The four roots of $f_1(z) = 0$ are denoted by $z_1$, $z_2$, $z_3$ and $z_4$. According to Vieta theorem, we have the following relations,
\bea
z_1z_2z_3z_4=\frac{E_1}{A_1},\\
z_1z_2+z_1z_3+z_1z_4+z_2z_3+z_2z_4+z_3z_4=\frac{C_1}{A_1}.
\eea

Based on the asymptotic behaviors of the effective potential at the inner and outer horizons and spatial infinity, one can deduce that equation $V'(z)=0(\text{or}~ f_1(z)=0)$ has at least two positive real roots when $z> 0$. We denote these two positive real roots by $z_1, z_2$.

In the following, we will find a parameter region of the system where there are only  two positive roots for the equation $f_1(z)=0$, i.e., $z_3,z_4$ are both negative. This will be achieved by requiring
\bea
E_1>0,~~C_1<0,
\eea
under the condition \eqref{mu1}. Then there is no trapping potential well acting as a mirror outside the outer horizon of the Kerr-Newman black hole. The black hole and scalar perturbation system is superradiantly stable in this region.

\subsection{Analysis of the coefficient $E_1$}
The coefficient $E_1$ is lengthy and involves a separation constant $\lambda_{lm}$ without exact analytic expression, so it seems that it is not easy to get a general analytic proof for its sign. However, we find that this is not the case. An expectedly simple proof for the sign of $E_1$ will be provided in the following of this section.

Take $E_1$ as a quadratic function of $\omega$, which is written as follows,
\begin{equation}
E_1\left( \omega \right) =a_{E_1}\omega ^2+b_{E_1}\omega +c_{E_1},
\end{equation}
where
\begin{equation}
\begin{split}
a_{E_1}&=4( a^2+r_{-}^{2})[ a^2M+r_-Q^2-Mr_{-}^{2}],
\\
b_{E_1}&=-2a( aqQ^3-2mMQ^2+a^3qQ+2a^2mM)\\
&\quad -2r_-[ 2a( 2mQ^2+aMqQ+2mM^2)\\
&\quad -r_-( qQr_{-}^{2}+2qQMr_-+6amM-3qQ^3)] ,
\\
c_{E_1}&=4M^3+2a^3mqQ+2amqQ^3-2a^4M\mu ^2\\
&\quad -2MQ^2( 2+\lambda _{lm} ) -2a^2M( 2-2m^2+Q^2\mu ^2\\
&\quad +\lambda _{lm} ) +4amMqQr_--2a^2r_-[ -2+2m^2\\
&\quad -q^2Q^2+( -2M^2+Q^2) \mu ^2-\lambda _{lm} ] \\
&\quad +4M^2r_-( -1+\lambda _{lm}) +2Q^2r_-[ 2\\
&\quad +Q^2( q+\mu )( q-\mu) +\lambda _{lm}] -2r_{-}^{2}\{ 3[ amqQ\\
&\quad +M( -Q^2\mu ^2+\lambda _{lm})] +r_-[ 2M^2\mu ^2\\
&\quad +Q^2( q^2+\mu ^2) -\lambda _{lm}-M\mu ^2r_-]\}.
\\
\end{split}
\end{equation}

Firstly, the coefficient $a_{E_1}$ can be simpified by the relationship between $M$, $a$, $Q$ and $r_-$, $r_+$ as follows,
\begin{equation}
a_{E_1}=2\left( a^2+r_{-}^{2} \right) ^2\left( r_+-r_- \right) ,
\end{equation}
which is obviously positive, i.e.,$a_{E_1}>0$. This means $E_1$ will be larger than $0$ for all $\omega$ if $\varDelta _{E_1}<0$. The discriminant of $E_1$ can be calculated directly as follows,
\begin{equation}
\begin{split}
\varDelta _{E_1}&=64(a^2 - M^2 + Q^2)^2(4a^2M^2 - 8M^4 \\
&\quad + 8M^2Q^2- Q^4 + 8M^3 \sqrt{-a^2 + M^2 - Q^2} \\
&\quad - 4MQ^2 \sqrt{-a^2 + M^2 - Q^2})\\
&\quad =-64(a^2 - M^2 + Q^2)^2( a^2+r_{-}^{2}) ^2<0.
\end{split}
\end{equation}
So we conclude that for any $\omega$
\bea
E_1>0.
\eea
This is an interesting result. It is noteworthy to point out  that the separation constant $\lambda_{lm}$ disappears in the discriminant $\varDelta _{E_1}$.

 Then according to the analysis above, we have
\begin{equation}
z_1 z_2 z_3 z_4>0.
\end{equation}
As we mentioned before, $z_1$ and $z_2$ are two positive roots of the derivative of the effective potential. From the above equation, we find that $z_3$ and $z_4$ can only be both positive or both negative when they are real roots.

\subsection{Analysis of the coefficient $C_1$}
In this section, we want to identify a parameter region where $C_1<0$, then the roots $z_3, z_4$ are both negative under the condition \eqref{mu1}.
The expression of $C_1$ is
\bea\label{C1111}\nonumber
C_1&=&-3(r_+-r_-)\lambda_{lm}+12r_-(Q^2-r_-^2-r_+r_-)\omega^2\\\nonumber
&&+6[am(r_-+r_+)-(Q^2-3r_-^2-r_-r_+)qQ]\omega\\\nonumber
&&-3(Q^2r_--Q^2r_+-r_-^3+r_+^2 r_-)\mu^2\\
&&-6amqQ-6q^2Q^2r_-.
\eea
Here, for the eigenvalue of the spheroidal angular equation $\lambda _{lm}$, we choose the following lower bound \cite{Hod:2016iri},
\bea
\lambda _{lm}>m^2 -a^2( \mu ^2-\omega ^2).
\eea
Due to the negative coefficient of $\lambda_{lm}$ in \eqref{C1111}, we can obtain following inequality,
\begin{equation}
\begin{split}
C_1&<-3(r_+-r_-)[m^2 -a^2( \mu ^2-\omega ^2)]\\
&\quad+12r_-(Q^2-r_-^2-r_+r_-)\omega^2\\
&\quad+6[am(r_-+r_+)-(Q^2-3r_-^2-r_-r_+)qQ]\omega\\
&\quad-3(Q^2r_--Q^2r_+-r_-^3+r_+^2 r_-)\mu^2\\
&\quad-6amqQ-6q^2Q^2r_-\\
&\quad=-3r_-^2(r_+-r_-)\mu^2\\
&\quad+3[2(a^2+2Q^2)r_--(a^2+4r_-^2)(r_-+r_+)]\omega^2\\
&\quad+6[-qQ^3+qr_-(3r_-+r_+)Q+am(r_-+r_+)]\omega\\
&\quad-6q^2r_-Q^2-6amqQ+3m^2(r_--r_+).
\end{split}
\end{equation}
Considering the inequality \eqref{bsc} and the negative coefficient of $\mu$, the above inequality can be further written as
\bea\nonumber
C_1&<&-3(3r_-+r_+)(a^2+r_-^2)\omega^2\\\nonumber
&&+6[-qQ^3+qr_-(3r_-+r_+)Q+am(r_-+r_+)]\omega\\
&&-6q^2r_-Q^2-6amqQ+3m^2(r_--r_+).
\eea

Now we treat the right side of the above inequality as a quadratic function of $\omega$,
\bea\nonumber
g(\omega)&=&-3(3r_-+r_+)(a^2+r_-^2)\omega^2\\\nonumber
&&+6[-qQ^3+qr_-(3r_-+r_+)Q+am(r_-+r_+)]\omega\\
&&-6q^2r_-Q^2-6amqQ+3m^2(r_--r_+).
\eea
Identifying a parameter region where $g(\omega)<0$ will be a sufficient condition for $C_1<0$ in the same region. We will do this in the following.

It is easy to see that the quadratic coefficient of $g(\omega)$ is negative. So if the determinant of $g(\omega)$, $\Delta_{g(\omega)}$, is negative, then $g(\omega)<0$.
Here we take $\Delta_{g(\omega)}$ as a quadratic function of the  azimuthal quantum number $m$, which is expressed as
\bea
\Delta_{g(\omega)}\equiv K(m)=36[a_{K(m)}m ^2+b_{K(m)}m +c_{K(m)}],
\eea
where
\bea\nonumber
a_{K(m)}&=&r_{-}^{2}(3r_{-}^{2}+2r_+r_--r_{+}^{2}-4Q^2),\\\nonumber
b_{K(m)}&=&4aqQ^3r_- ,\\
c_{K(m)}&=&q^2Q^2(3r_{-}^{4}-2r_+r_{-}^{3}-r_{+}^{2}r_{-}^{2}+Q^4).
\eea

First, let's consider the quadratic coefficient of $K(m)$, $a_{K(m)}$. Because $Q^2>0$, $a_{K(m)}$ will be negative if the following inequality holds,
\bea
3r_{-}^{2}+2r_+r_--r_{+}^{2}\leqslant0,
\eea
i.e.
\begin{equation}\label{1/3}
\frac{r_-}{r_+}\leqslant \frac{1}{3}.
\end{equation}
This is another important result in this paper.

Next, we consider the discriminant of $K(m)$, $\Delta_{K(m)}$. The expression of $\Delta_{K(m)}$ is
\begin{equation}
\begin{split}
\Delta_{K(m)}&=4q^2Q^2r_{-}^{2}(r_{-}^{2}+a^2)[r_-(3r_{+}-r_-)+Q^2]\\
&\quad\times(3r_{-}^{2}-2r_+r_--r_{+}^{2}).
\end{split}
\end{equation}
On the right side of the above equation, factors are all obviously positive except the last factor.
With the relation $\frac{r_-}{r_+}\leqslant \frac{1}{3}$ in \eqref{1/3}, we can easily prove the sign of the last factor is negative.
\begin{equation}
\begin{split}
3r_{-}^{2}-2r_+r_--r_{+}^{2}=r_{+}^{2}[3(\frac{r_-}{r_+})^2-2(\frac{r_-}{r_+})-1]<0.
\end{split}
\end{equation}
Then,
\begin{equation}
\Delta_{K(m)}<0.
\end{equation}
So under the condition $\frac{r_-}{r_+}\leqslant \frac{1}{3}$,
\begin{equation}
\begin{split}
\Delta_{g(\omega)}=K(m)<0.
\end{split}
\end{equation}
Then we finally have
\bea
g(\omega)<0,~~C_1<g(\omega)<0.
\eea
The two real roots $z_3,z_4$ should be both negative. Then there is only a maximum outside the outer horizon  for the effective potential. No trapping potential well exists from outer horizon to spatial infinity.

\section{Summary}
In this work, the superradiant stability property is analytically studied for a system consisting of a background Kerr-Newman black hole and a charged massive scalar perturbation. The equation of motion of the minimally coupled scalar perturbation in the KN black hole background is separated into angular and radial parts. In our study, the spheroidal angular equation is prolate. The radial equation of motion of the incoming scalar field is written as a Schrodinger-like equation and the superradiant stable parameter space region of the system is determined by analyzing the effective potential $V(r)$.

Under the conditions that the superradiant modes exist in the KN black hole and scalar perturbation system, if there is no trapping potential well outside the KN black hole outer horizon $r_+$, the system is superradiantly stable. By analysing the derivative of the effective potential $V(r)$, it is found that when the ratio of inner and outer horizons of the KN black hole satisfies
\bea
\frac{r_-}{r_+}\leqslant \frac{1}{3},
\eea
and the charge-to-mass ratios of the KN black hole and scalar perturbation satisfy
\bea
\frac{q}{\mu}\frac{Q}{M}>1,
\eea
the KN black hole is superradiantly stable against charged massive scalar perturbation.

{\bf Acknowledgements:}\\
We would like to thank Lei Yin and Zhan-Feng Mai for helpful discussion. This work is partially supported by  Science and Technology Program of Guangzhou (No. 2019050001)
 and Natural Science Foundation of Guangdong Province (No. 2020A1515010388).

\end{document}